# Thermal Analysis and SANS Characterization of Hybrid Materials for Biomedical applications

**J. J. H. Lancastre · F. M. A. Margaça · L. M. Ferreira · A. N. Falcão**

**I. M. Miranda Salvado · M. S. M. S. Nabiça · M. H. V. Fernandes**

**L. Almásy**

---

J. J. H. Lancastre (✉) · F. M. A. Margaça · L. M. Ferreira · A. N. Falcão
Physics and Accelerators Unit, Nuclear and Technological Institute,
E.N. 10, 2686-953 Sacavém, Portugal
Telephone: +351 21 994 6045
Fax: +351 21 994 1525
e-mail: jlancastre@itn.pt

M. Miranda Salvado · M. S. M. S. Nabiça · M. H. V. Fernandes
Department of Glass and Ceramics Engineering, CICECO,
Aveiro University, 3810-193 Aveiro, Portugal

L. Almásy
Laboratory for Neutron Scattering, PSI, CH-5232 Villigen, Switzerland
Research Institute for Solid State Physics and Optics**,** POB 49, 1525-Budapest, Hungary

**Abstract**

Silicate hybrid materials were prepared by the sol-gel process with the addition of $x$ wt% of zirconium propoxide ($x$ = 0 and 1). The thermal behavior as well as the influence of Zr addition was studied by thermal gravimetric analysis and differential thermal analysis. The microstructure evolution with temperature was investigated by X-ray diffraction and small angle neutron scattering. It was found that the beginning of polymer degradation occurs at a higher temperature in the material prepared with addition of Zr than in the one prepared without. At the nanometric scale, the materials prepared without Zr show smooth interfaces, whereas those with Zr present a mass fractal structure. This structure is also observed in the material without Zr after thermal treatment at 200 ºC. The results showed that bioactivity is favored by mass fractal structures in comparison with one consisting of smooth surfaces.

**Introduction**

Sintered hydroxyapatite [1], $Na_2O$-$CaO$-$SiO_2$-$P_2O_5$ glasses [2] and glass-ceramics containing crystalline apatite and wollastonite [3] can bond to living bone and are currently applied as important bone repairing materials [4]. Metals, such as titanium and its alloys, can be also used as living bone bonding materials [5]. However metals present larger elastic modulus than the human cortical bones and in this way they can induce resorption of the surrounding bone due to stress shielding. Therefore there is the need for the preparation of bioactive materials with lower elastic modulus. In the last decade organically modified sol-gel materials such as Ormosils based on the mixture of tetraethylortosilicate (TEOS) and polydimethylsiloxane (PDMS) have been the focus of attention due to their unique properties such as high flexibility and low elastic modulus together with high mechanical strength [6]. In these materials, also known as hybrid materials, the organic polymer is chemically incorporated at molecular level into the inorganic network [7, 8].

Several authors prepared PDMS/TEOS based materials for use as bioactive materials: Chen et al. [6] refer the preparation of bioactive PDMS-modified $CaO$-$SiO_2$-$TiO_2$ hybrids and Tsuru et al. [9] reported the preparation of PDMS-modified $CaO$-$SiO_2$ hybrids that showed apatite forming ability after immersion in a simulated body fluid (SBF). Also Yabuta et al. [10] developed PDMS-based porous materials with the ability of apatite deposition on the pore walls after 3 days of immersion in SBF.

The authors prepared hybrid materials PDMS-modified $CaO$-$SiO_2$-$ZrO_2$ and studied the bioactivity behaviour after immersion in SBF for different periods of time. It was found that both the addition of $ZrO_2$ and the thermal treatment were beneficial for the deposition of a hydroxyapatite surface layer on the synthesised materials [11]. In order to clarify the roles of Zr and the thermal treatment, the thermal behaviour of the prepared materials, as well as their microstructure, were studied by thermal analysis, X-ray diffraction and small angle neutron scattering (SANS) in the present work.

**Experimental**

Polydimethylsiloxane (PDMS), S12, with average molecular weight 400-700 , ABCR GmbH & Co., tetraethylortosilicate (TEOS), Merck, ethanol absolute, Merck, calcium nitrate tetrahydrate, Merck, zirconium propoxide (PrZr), Fluka and ethylacetoacetate (Eacac), Merck, were used to prepare the hybrid materials with the composition, in weight %, shown in Table 1. For the preparation of the hybrid materials the appropriate amount of TEOS was mixed with ethanol absolute in a molar ratio equal to 3.5. To this solution 2.50 g of PDMS together with 0.017 moles of ethylacetoacetate and later a solution of calcium nitrate tetrahydrate (molar ratio $H_2O$/Calcium nitrate equal to 13) were added respectively. This mixture is here referred as solution A. In the case of samples with zirconium, PrZr mixed with Eacac in a molar ratio Eacac/PrZr equal to 1.0 was added to the solution A. All samples were dried at 120 ºC for one week. Some of these were then heat treated at 200ºC for 2 hours.

**Table 1** Preparation conditions of the materials. All samples were dried at 120 ºC for one week. Afterwards, two samples were heat treated at 200 ºC during 2 hours

| Sample | PDMS (wt%) | TEOS (wt%) | $Ca(NO_3)_2$ (wt%) | PrZr (wt%) | Temperature (ºC) |
|---|---|---|---|---|---|
| **T120-Zr0** | 17 | 71 | 12 | 0 | 120 |
| **T120-Zr1** | 17 | 70 | 12 | 1 | 120 |
| **T200-Zr0** | 17 | 71 | 12 | 0 | 200 |
| **T200-Zr1** | 17 | 70 | 12 | 1 | 200 |

X-ray diffraction was performed using a Rigaku Geigerflex equipment, with CuKα (λ=1.5418 Å), and thermal analysis was carried out on dried samples using a Setaram equipment, model LAB SYS, at Aveiro University. The SANS measurements were performed using the SANS II instrument at the Paul Scherrer Institute, Villigen, Switzerland [12].

The neutron scattering experiments were performed using two different experimental settings (sample to detector distance of 6 m and 2 m with wavelength of 9.03 Å and 5.09 Å, respectively) to cover a broad range of scattering vectors $Q$ (from 0.004 Å$^{-1}$ to 0.2 Å$^{-1}$). The samples were measured at room temperature. Scattered intensities have been normalised for sample thickness and transmission, and corrected for detector efficiency by the use of an incoherent scatterer.

**Results and Discussion**

Thermal Analysis

The thermal study of the materials was carried out by thermogravimetric analysis (TGA) and differential thermal analysis (DTA). The results are shown in Figures 1 and 2, respectively.

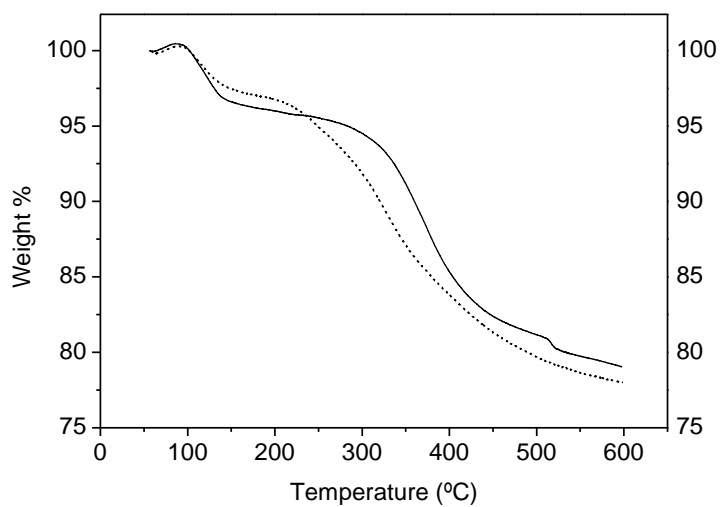

**Fig. 1** TGA curves of samples T120-Zr0 (dashed line) and T120-Zr1 (solid line)

It is observed that the sample prepared with Zr, T120-Zr1, is more stable with the temperature increase until the beginning of polymer degradation that occurs approx. at 320 ºC than the one prepared without Zr, T120-Zr0, that begins to degrade around 220 ºC. This result is in accordance with the thermal stabilizing effect of Zr observed by the authors [13] in hybrid materials with different composition of precursors and contents.

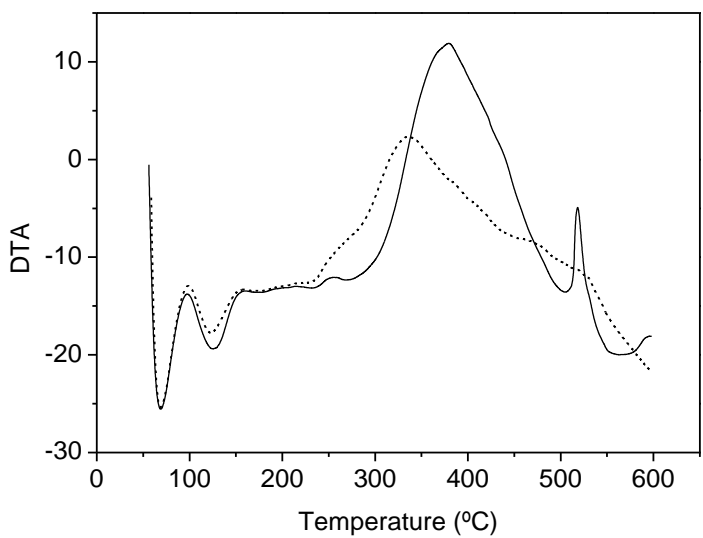

**Fig. 2** DTA curves of as dried samples T120-Zr0 (dashed line) and T120-Zr1 (solid line)

The DTA results show that both samples exhibit a small endothermic peak at approximately 130 ºC attributed to the loss of physically adsorbed water and a large exothermic broad peak related mainly to polymer degradation and consequent structural reorganization.

The degradation of the sample prepared without Zr, T120-Zr0, takes place at a lower temperature and proceeds at a slower rate than the one prepared with Zr. In this sample, T120-Zr1, the shape of the exothermic peak reveals that a larger amount of energy is released in a faster manner and at a higher temperature. This is attributed to a stronger matrix structure, developed due to the presence of Zr [13].

X-ray Diffraction Results

To proceed, the structure of the materials, at the atomic scale, was investigated by X-ray diffraction (XRD). The XRD spectra of the samples prepared without and with Zr, are shown in Figures 3 and 4, respectively.

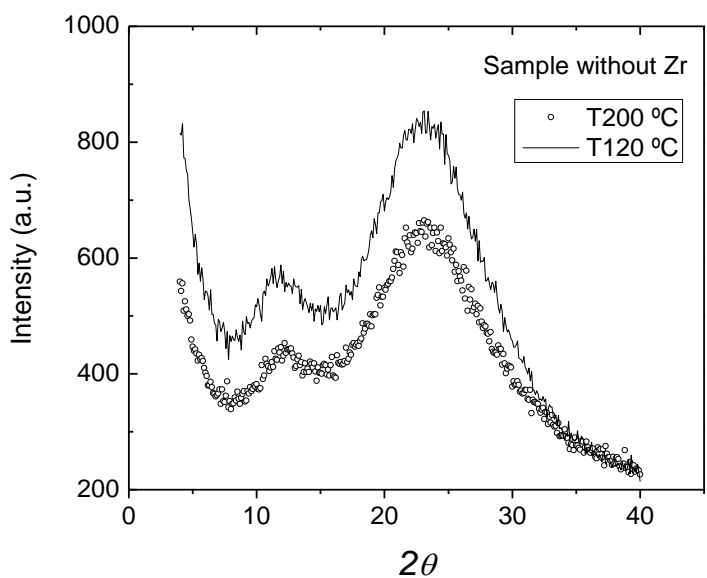

**Fig. 3** XRD curves obtained for samples prepared without Zr, as dried T120-Zr0 (solid line) and heat treated, T200-Zr0 (open circle)

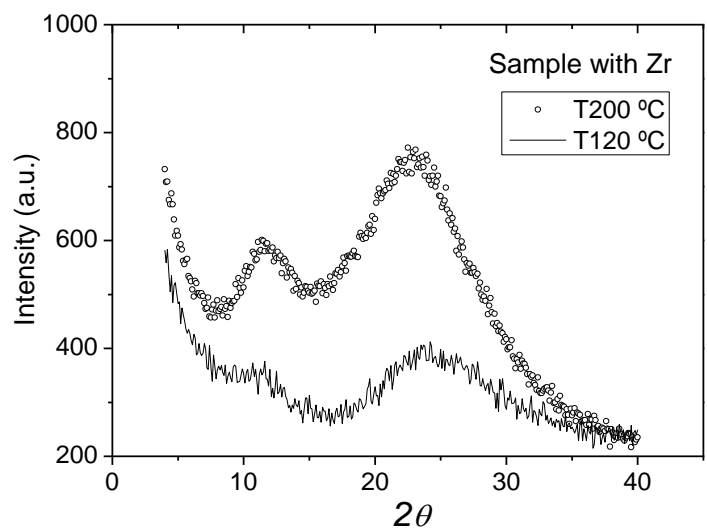

**Fig. 4** XRD curves obtained for samples prepared with Zr, as dried T120-Zr1 (solid line) and heat treated, T200-Zr1 (open circle)

The spectra show two broad peaks, one at 2θ ~ 11° and another at 2θ ~ 22°. The first peak corresponds to ca. 8 Å distances and is attributed to ordered segments of polymer chains in a folded configuration [14]. The second peak is characteristic of amorphous silica.

In the case of samples without Zr, as shown in Fig. 3, the contributions of ordered polymer chain segments and amorphous silica decrease significantly upon heat treatment at 200 ºC. This was expected due to the beginning of polymer degradation as indicated before by thermal analysis results. On the other hand, samples prepared with Zr, after heat-treatment at 200 ºC, show an increase in intensity both for the amorphous silica peak and for that associated to the ordered segments of polymer chains. This is in accordance with the stabilizing effect of Zr in this type of hybrid materials, which in consequence leads to the increase in its thermal stability [13]. Furthermore this can be attributed to a network modifying behaviour of Zr in these materials, mentioned previously by other authors to explain the high catalytic activity of Zr [15].

SANS Results

Small-angle scattering provides structural information in the size range ca. 1-100 nm. The neutron scattering in these materials is due to the contrast between the inorganic oxide matrix and the empty pores or PDMS chains present in its midst. Considering Si oxide as the most abundant in the inorganic matrix, the coherent scattering length densities, $\rho$, are estimated as ~ $3 \times 10^{10}$ cm$^{-2}$ for the matrix, as 0 cm$^{-2}$ for pores and as ~ $0.067 \times 10^{10}$ cm$^{-2}$ for PDMS. The SANS contrast between the scattering object ($\rho$) and the matrix ($\rho_0$), defined as ($\rho - \rho_0$), is approximately the same for both the empty pores and the PDMS, ca. $3. \times 10^{10}$ cm$^{-2}$. Therefore, for this material the scattering from pores or from PDMS is undistinguishable.

First, consider the materials prepared with Zr. Figure 5 shows SANS spectra of materials prepared with 1 wt% PrZr, as dried and after thermal treatment.

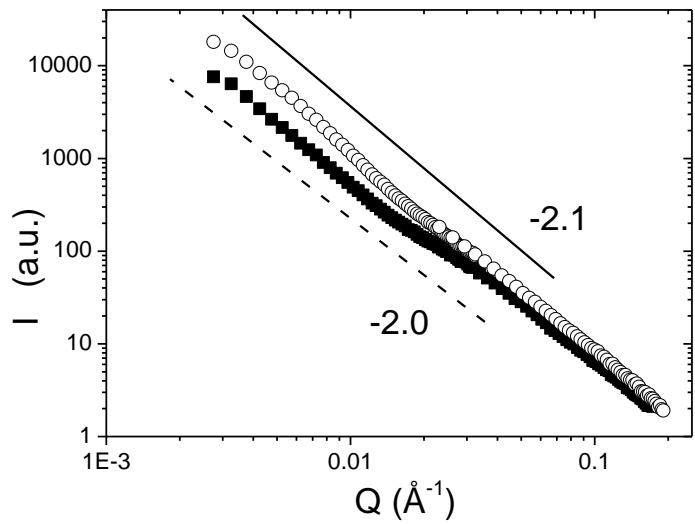

**Fig. 5** SANS spectra of samples T120-Zr1 (full circle) and T200-Zr1 (open circle). The lines corresponds to slopes -2.0 and -2.1

The scattering from the sample T120-Zr1 is dominated by a linear descent of slope ca. - 2.0 for the whole Q range covered in the measurement. The SANS pattern obtained for the heat treated sample T200-Zr1 is similar except for the slope value that is ca. - 2.1. Both these samples show an open network with mass fractal structure. The slightly higher fractal density for the heat-treated sample reveals the strengthening of the network induced by the thermal treatment, as predicted by the DTA curve. Thus the material shows a matrix with mass fractal porosity.

Figure 6 shows the SANS spectra for materials prepared with no addition of PrZr, before and after thermal treatment.

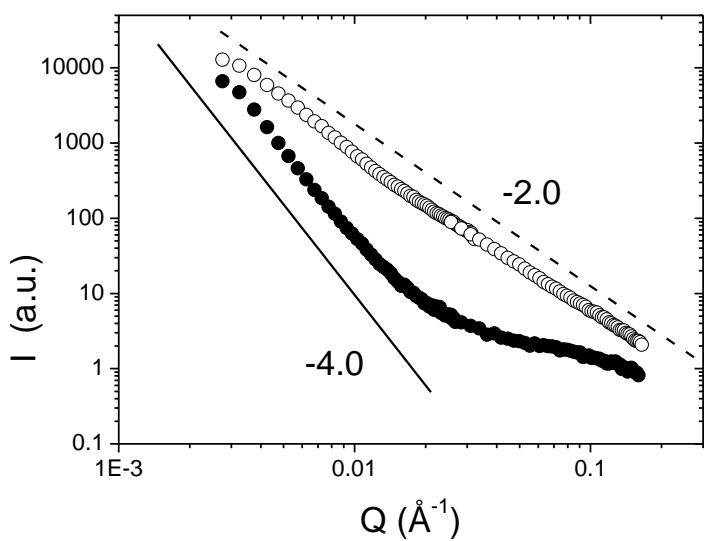

**Fig. 6** SANS spectra of samples T120-Zr0 (full circle) and T200-Zr0 (open circle). The lines correspond to slopes -2 and -4

The different shapes for the scattering curves obtained for the dried and heat-treated samples prepared without Zr reveal that the structure changed drastically upon the heat treatment. This was expected from the observed TGA and DTA curves obtained for these materials.

The dried sample exhibits Porod behaviour (I ~ $Q^{-4}$), at Q < 0.015 Å$^{-1}$ indicating a two-phase structure with a characteristic size 2π/0.015 ~ 400 Å, with smooth interfaces. This might be either a porous structure with pore of ca 40 nm or a possible segregation of PDMS from the silica matrix as SANS can not distinguish the true pores from polymer regions. In this case the system consists of large aggregated structures of small unites that are the low molecular weight polymer chains. The surface of the aggregates obtained by drying at 120 °C is smooth [16].

The sample heat treated at 200 °C for 2 hours, shows a linear descent with slope of –2, which according to Schaefer et al. cited by Brinker and Scherer [17], is expected from an open inorganic oxide network with mass fractal structure. The thermal analysis results of this material showed that at this temperature the polymer begins to degrade besides the release of residual organic species resulting from the hydrolysis of the precursors. The combination of both processes leads to formation of an oxide network of the mass fractal type structure.

The samples without Zr as dried and after heat treatment gave different results in bioactivity tests [11]. The precipitation of apatite was favoured for the sample T200-Zr0 that has mass fractal structure in comparison with the sample T120-Zr0 which presents smooth surfaces. This is similar to the results obtained by other authors that have found a correlation between surface roughness and the apatite precipitation on different materials after soaking in SBF [18, 19].

Consider how these different structures can develop in the sol. It is well known that in sol-gel synthesis, under most conditions, the obtained materials at the nanometer length scale possess tenuous structures which can be well described as mass or surface fractals [17]. Several models have been proposed [20] to describe the network growth in silicate systems. It was found that fractal structures generally emerge unless the growth proceeds predominantly by reaction-limited addiction of monomers to a growing cluster (a process similar to the classical nucleation and growth). In silicates, the condensation rate constant is sufficiently small, so reaction-limited conditions dominate, leading to compact structures with smooth surfaces that give the Porod slope - 4 in small angle scattering measurements [21]. This growth model is known as the Reaction Limited Monomer-Cluster Aggregation (RLMCA). It describes the development of the microstructure of the sample T120-Zr0. Thermal treatment of materials can change significantly the structure, depending on the temperature and the presence of Zr.

When Zr is present in the precursor's mixture the chemistry changes significantly, in as much as Zr is known to have a catalyst effect in the sol-gel processing of silicate species [22]. The presence of Zr favours Reaction Limited Cluster-Cluster Aggregation (RLCCA). This process leads to a mass fractal structure, with the fractal dimension of 2.09 [17]. This is in fairly good agreement with the slope found for sample Zr1T120.

Finally, Figure 7 presents SEM micrographs previously obtained [11] that show a few of the studied samples after being immersed in SBF. According to the findings of the bioactivity tests [11] the precipitation of apatite was favoured by the presence of Zr. The SEM micrographs show that the most abundant and homogeneous coating of apatite is deposited in the material that was prepared with Zr addition and heat-treated before immersion, sample T200Zr1. The identification of the deposited layer as hydroxyapatite was confirmed by EDS and XRD analysis [11]. These results together with those of SANS further relate the enhanced precipitation to the mass fractal type structure.

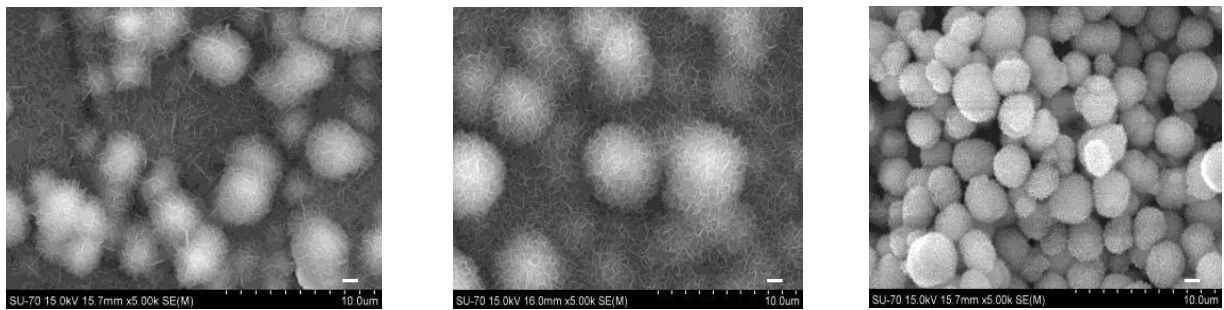

**Fig. 7** SEM micrographs of T120Zr0 (left), T120Zr1 (middle) and T200Zr1 (right) after being 14 days immersed in SBF. The white scale bar measures 1 μm

The bioactivity results showed that the enhanced precipitation is related to the presence of Zr and now the SANS results indicate that the nanoscale structure of the prepared material is directly related to its bioactivity. Since the structure changes with the presence of Zr further work is necessary to ascertain whether the enhanced precipitation is due to the presence of Zr or to the change of the structure or to both.

**Conclusions**

Materials of the system PDMS-CaO-$ZrO_2$-$SiO_2$, with and without Zr, have been prepared by the sol-gel process and characterised by TGA, DTA, XRD and SANS. The results showed that in the material prepared without Zr the oxide network evolved by RLMC aggregation, resulting in smooth interfaces at the nanometric scale. On the other hand, the oxide network of samples with PrZr grew by RLCC aggregation presenting a mass fractal structure, at the nanometric scale. The observed correlation between nanoscale structure and bioactivity should be further investigated in order to improve the processing conditions of materials for biomedical applications.

## Acknowledgements

The work was based on experiments performed at the Swiss spallation neutron source SINQ, Paul Scherrer Institute, Villigen, Switzerland, and it was supported by the FCT (Portuguese Foundation for Science and Technology).

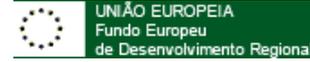